\documentclass[
aps,prd,
12pt,
nopreprintnumbers,
showpacs,
eqsecnum,
nofootinbib
]{revtex4-1}

\usepackage{graphicx}

\def\vx{{\bf x}}
\def\vr{{\bf r}}
\def\vk{{\bf k}}
\def\vl{{\bf l}}

\def\vq{{\bf q}}
\def\vp{{\bf p}}

\def\v0{{\bf 0}}

\begin{document}

\title{
Interparticle Potential up to Next-to-leading Order 
for Gravitational,
Electrical, and Dilatonic Forces
}
\author{Nahomi Kan}\email[]{kan@yamaguchi-jc.ac.jp}
\affiliation{
Yamaguchi Junior College,
Hofu-shi, Yamaguchi 747--1232, Japan}
\author{Kiyoshi Shiraishi}\email[]{shiraish@yamaguchi-u.ac.jp}
\affiliation{
Yamaguchi University,
Yamaguchi-shi, Yamaguchi 753--8512, Japan}
\date{\today}

\begin{abstract}
Long-range forces up to next-to-leading order are computed in the
framework of the Einstein-Maxwell-dilaton system by means of a
semiclassical approach to gravity. As has been recently shown, this
approach is effective
if one of the masses under consideration is significantly greater than all
the energies involved in the system. Further, we obtain the condition for
the equilibrium of charged masses in the system.
\end{abstract}



\maketitle

\section{Introduction}
In Newtonian dynamics, the interaction between two charged massive
particles with masses and charges of $M$, $m$ and $Q$, $q$,
respectively, is described by the Newton and Coulomb potentials that
depend only on the distance between the two particles, $r$,
\begin{equation}
V(r)=-G\frac{Mm}{r}+\frac{1}{4\pi}\frac{Qq}{r},
\label{NC}
\end{equation}
where $G$ denotes the Newton constant.
If $Qq=4\pi GMm$, the long-range forces cancel each other out. 
The static exact general-relativistic solution for the equilibrium of two
or more charged masses was obtained by
Majumdar and Papapetrou~\cite{MP}.
The Majumdar-Papapetrou solution is given as%
\footnote{We use units such that $c=1$ and $\hbar=1$ throughout the
present paper.}
\begin{equation}
ds^2=V^{-2}dt^2-V^2 d\vx^2\,,
\end{equation}
with
\begin{equation}
V=1+\sum_i\frac{GM_i}{|\vr-\vr_i|}\,,
\end{equation}
and the electric potential
\begin{equation}
A_0=1-V^{-1}\,.
\end{equation}
Here $\vr=(x,y,z)$ and the $i$-th charged particle is located at $\vr_i$.
The charge of each particle can be read as
$Q_i=\sqrt{ 4 \pi G}M_i$.
The nonlinearity in general relativity means that 
the possible higher-order interactions other than that given
by Eq.~(\ref{NC}), are canceled if the critical mass-charge relation is
fulfilled.

Most recent modified gravity theories contain scalar
fields as elementary or effective degrees of freedom for mediating an
additional force to the Einsteinian/Newtonian gravitational force.
Thus, the incorporation of scalar forces in the interaction of many-body
systems is of considerable interest in various contexts of particle
physics and theoretical astrophysics.

In classical theory, the possible cancellation of three long-range forces
has been proposed and discussed. The static exact
solution with a dilaton field was obtained by one of the present
authors~\cite{KSJMP}.  The Lagrangian for the fields that mediate
long-range forces is, in this case,
\begin{equation}
{\cal
L}=\frac{\sqrt{-g}}{4}\left(R-e^{-2a\phi}g^{\mu\rho}g^{\nu\sigma}
F_{\mu\nu}F_{\rho\sigma}+2g^{\mu\nu}\partial_\mu\phi\partial_\nu\phi
\right)\,,
\label{EMD}
\end{equation}
where $R$ denotes the scalar curvature derived from the metric
$g_{\mu\nu}$ and $F_{\mu\nu}\equiv\partial_\mu A_\nu-\partial_\nu A_\mu$
as usual. The dilaton field is denoted by $\phi$, and $a$ denotes the
dilaton coupling constant.
For simplicity, we set $4\pi G=1$.
The metric for the solution is written by
\begin{equation}
ds^2=U^{-2}dt^2-U^2 d\vx^2\,,
\label{Smetric}
\end{equation}
with
\begin{equation}
U=V^{\frac{1}{1+a^2}}\,,
\end{equation}
and
\begin{equation}
V=1+\sum_i\frac{(1+a^2)M_i}{4\pi|\vr-\vr_i|}\,,
\end{equation}
while the electrostatic potential and the dilaton field are given by
\begin{equation}
A_0=\frac{1}{\sqrt{1+a^2}}(1-V^{-1})\,,\qquad
e^{-2a\phi}=V^{\frac{2a^2}{1+a^2}}\,.
\label{Ssol2}
\end{equation}
The solution is valid for the case with the
balance condition
$(M_i:Q_i:\Sigma_i)=(1:\sqrt{1+a^2}:a)$, 
where $\Sigma$ denotes the dilatonic charge.

In the present paper,
we calculate the next-to-leading potential in the
Einstein-Maxwell-dilaton system using the Feynman diagram technique.
We verify the cancellation of long-range forces in the
Einstein-Maxwell-dilaton system under the
balance condition, $Q_i=\sqrt{1+a^2}M_i$.

We use the method of perturbative quantum field theory to obtain the
effective potential for the two-body problem of charged
sources~\cite{Paszko}. 
The advantage of the method is that we can extend the analysis of
interactions to the one including quantum effects in a straightforward
manner in future.\footnote{The loop correction to the potential of
electrically charged masses was evaluated by many
authors~\cite{BB,Faller}.} The method can also be extended in another
direction, that is, the $n$-body problem can be investigated by the
perturbative method~\cite{Chu}. Another advantage of the
perturbative method is that we can extract and investigate a necessary
contribution only, snd this can lead to illuminative
discussions regarding the complex nature of such interacting many-body
systems.


This paper is organised as follows. In Sec.~\ref{source}, we introduce the
Lagrangian for a charged scalar field as a source of long-range forces.
In Sec.~\ref{dilaton}, we introduce the perturbative tools for a dilaton
field. We use several Feynman diagrams to obtain the resulting potential
when three force-mediating fields exist.

Sec.~\ref{precession}, Sec.~\ref{external} and
Sec.~\ref{Hamiltonian} are devoted to the applications of the potential
obtained in Sec.~\ref{dilaton}. In Sec.~\ref{precession}, we show the
precession of the orbit of a charged dilatonic body. The correspondence
between the exact solution and the perturbative result is examined in
Sec.~\ref{external}. In Sec.~\ref{Hamiltonian}, we consider the case where
the static forces cancel each other; it is observed that the known
description of charged bodies with low velocities is reproduced.

We summarize our results in the last section and we provide an overview
of our study.

\section{Source and force fields
\label{source}}

We begin with the Lagrangian (\ref{EMD}) for force-mediating fields.
In order to 
evaluate the potential of long-range forces from Feynman diagrams,
we use a complex scalar field $\varphi$ as a source field, or in other
words, as a probe. The Lagrangian of the complex Klein-Gordon field
is~\cite{DS}
\begin{equation}
\mathcal{L}_{KG} = {\sqrt{-g}}\,[e^{-a\phi}g^{\mu\nu}(D_\mu\varphi)^*
D_\nu\varphi-m^2e^{a\phi}\varphi^*\varphi]~,  
\end{equation}
where 
$D_\mu\varphi=\partial_\mu\varphi+iqA_\mu\varphi$,%
\footnote{The manner of coupling with the dilaton field is not unique.
Please see \cite{DS}. We adopt the simplest form in the dilaton
coupling.}
$q$ denotes the electric charge of a scalar boson and $m$ denotes its
mass.

To treat the interactions perturbatively,  
we decompose the metric $g_{\mu\nu}$ 
into the flat background field $\eta_{\mu\nu}$
and the graviton field $h_{\mu\nu}$ as
\begin{equation}
g_{\mu\nu}=\eta_{\mu\nu}+\kappa h_{\mu\nu},
\end{equation}
where $\eta_{\mu\nu}=diag. (1,-1,-1,-1)$.
The coefficient is chosen as $\kappa=\sqrt{32\pi G}$, as used popularly
in many studies. According to our convention in this paper, {\it i.e.},
$4\pi G=1$, we obtain $\kappa=\sqrt{8}$; nevertheless we continue to use
$\kappa$ as long as it does not cause confusion.

In this decomposition, the inverse of the metric becomes
\begin{equation}
g^{\mu\nu}=\eta^{\mu\nu}-\kappa h^{\mu\nu}
+\kappa^2 h^{\mu\lambda}h^\nu_\lambda
-\kappa^3 h^{\mu\lambda}h_{\lambda\alpha}h^{\alpha\nu}+\cdots~,
\end{equation}
and the square-root of the determinant of the metric is written as
\begin{equation}
\sqrt{-g}=\sqrt{-\det{g_{\mu\nu}}}
=1+\frac{\kappa}{2}h
+\frac{\kappa^2}{8}(h^2-2h^{\mu\nu}h_{\mu\nu})
+\frac{\kappa^3}{48}(h^3-6hh^{\mu\nu}h_{\mu\nu}+8h^\mu_\nu h^\nu_\lambda h^\lambda_\mu)+\cdots~,
\end{equation}
where $h^{\mu\nu}\equiv\eta^{\mu\alpha}h_{\alpha\beta}\eta^{\beta\nu}$ and
$h\equiv\eta^{\mu\nu}h_{\mu\nu}$.

Using these expansions, we obtain 
the Einstein-Hilbert action as follows: 
\begin{eqnarray}
\mathcal{L}_{EH}&=&\frac{1}{16\pi G}\sqrt{-g}R=
\frac{2}{\kappa^2}\sqrt{-g}R\nonumber \\
&=&\frac{1}{2}\left(\partial^\mu h^{\nu\lambda}\partial_\mu h_{\nu\lambda}
-\frac{1}{2}\partial^\mu h\partial_\mu h\right)\nonumber\\
&&
+\kappa\left(\frac{1}{2}h^\alpha_\beta\partial^\mu h^\beta_\alpha\partial_\mu h
-\frac{1}{2}h^\alpha_\beta\partial_\alpha h^\mu_\nu\partial^\beta h^\nu_\mu
-h^\alpha_\beta\partial_\mu h^\nu_\alpha\partial^\mu h^\beta_\nu\right.\nonumber\\
&&\!\!+\left.
\frac{1}{4}h\partial^\alpha h^\mu_\nu\partial_\alpha h^\nu_\mu
+h^\beta_\mu\partial_\nu h^\alpha_\beta\partial^\mu h^\nu_\alpha
-\frac{1}{8}h\partial^\mu h\partial_\mu h\right)+\ldots~,
\label{EinHil}
\end{eqnarray}
where we use the de Donder gauge, 
$\partial_\mu h^\mu_\nu=\frac{1}{2}\partial_\nu h$.
This expression involves a kinetic term as well as terms for an 
infinite number of interactions among gravitons.

The Lagrangian for the Maxwell theory coupled with gravitons becomes
\begin{eqnarray}
\mathcal{L}_M&=&
-\frac{\sqrt{-g}}{4}g^{\alpha\beta}g^{\mu\nu}F_{\alpha\mu}F_{\beta\nu}
\nonumber\\
&=&-\frac{1}{4}\eta^{\alpha\beta}\eta^{\mu\nu}F_{\alpha\mu}F_{\beta\nu}
-\frac{\kappa}{2}h^{\mu\nu}
\left[-\eta^{\alpha\beta}F_{\alpha\mu}F_{\beta\nu}-
\eta_{\mu\nu}\left(-\frac{1}{4}\eta^{\gamma\delta}\eta^{\lambda\sigma}
F_{\gamma\lambda}F_{\delta\sigma}\right)\right]\nonumber\\
&&+\frac{\kappa^2}{4}
\left[\frac{1}{2}(h^2-2h^{\mu\nu}h_{\mu\nu})
\left(-\frac{1}{4}\eta^{\gamma\delta}\eta^{\lambda\sigma}
F_{\gamma\lambda}F_{\delta\sigma}\right)\right.\nonumber\\&&
+\left.
F_{\alpha\beta}F_{\mu\nu}(hh^{\alpha\mu}\eta^{\beta\nu}
-2h^{\alpha\lambda}h^\mu_\lambda\eta^{\beta\nu}-h^{\alpha\mu}h^{\beta\nu})
\right]+\ldots\,,
\label{Maxwell}
\end{eqnarray}
where we choose the Lorenz gauge, $\partial^\mu A_\mu=0$.

In addition, we obtain the coupling between the massive scalar boson and
the dilaton as follows:
\begin{eqnarray}
\mathcal{L}_{KG}&=&
\mathcal{L}_{0}-a\phi(D_\mu\varphi)^*D^\mu\varphi
-a m^2\phi\varphi^*\varphi
-\frac{\kappa}{2}h^{\mu\nu}
\left(\partial_\mu\varphi^*\partial_\nu\varphi-\eta_{\mu\nu}\mathcal{L}_0\right)\nonumber\\
&&+\frac{\kappa^2}{2}\left[\frac{1}{4}(h^2-2h^{\mu\nu}h_{\mu\nu})\mathcal{L}_0
+\left(h^\mu_\lambda h^{\lambda\nu}-\frac{1}{2}hh^{\mu\nu}\right)
\partial_\mu\varphi^* \partial_\nu\varphi\right]
+\ldots ~,
\label{KG}
\end{eqnarray}
where ${\cal L}_0$ denotes the Lagrangian in the flat spacetime,
and it is given as follows:
\begin{equation}
\mathcal{L}_{0}\equiv
\frac{1}{2}(\eta^{\mu\nu}D_\mu\varphi^*D_\nu\varphi-m^2\varphi^*\varphi)\,.
\end{equation}

\section{Interaction mediated by dilaton field
\label{dilaton}}

\begin{figure}[ht]
\begin{center}
\includegraphics[width=5cm]{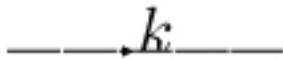}
\caption{Dilaton propagator.}
\label{dilatonpropagator}
\end{center}
\end{figure}

\begin{figure}[ht]
\begin{center}
\includegraphics[width=5cm]{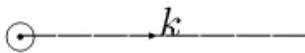}
\caption{External field of a dilaton with 
momentum ${\bf k}$.}
\label{def}
\end{center}
\end{figure}

Next, we consider a dilaton field.
The dilaton propagator (shown in Fig.~\ref{dilatonpropagator}) is
\begin{equation}
\frac{i}{k^2+i\epsilon}\,.
\end{equation}
We employ 
a dilatonic charge
$\rho_{\Sigma}=\Sigma\,\delta^{(3)}(\vx)$
as a classical source.
This source 
creates an external field $\phi^{ext}(\vk)$,
shown in Fig.~\ref{def}, which is given as follows:
\begin{equation}
\phi^{ext}(\vk)\equiv -\frac{\Sigma}{\vk^2}\equiv -\frac{aM}{\vk^2}~.
\end{equation}
The relation $\Sigma=aM$ is confirmed classically in the model described
by the same Lagrangian in Sec.~\ref{source}~\cite{DS}.

\begin{figure}[ht]
\begin{center}
\includegraphics[width=5cm]{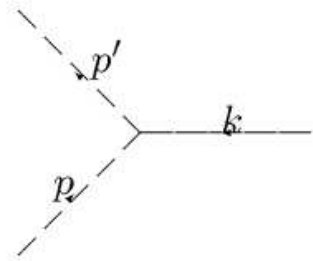}
\caption{Dilaton-scalar-scalar vertex.}
\label{dss}
\end{center}
\end{figure}

\begin{figure}[ht]
\begin{center}
\includegraphics[width=5cm]{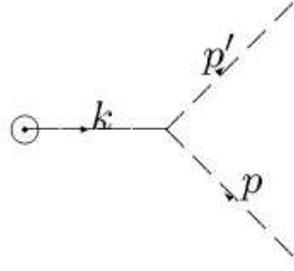}
\caption{First-order semiclassical amplitude with one dilaton,
$i\mathcal{M}_d^{(1)}$.}
\label{iMd1}
\end{center}
\end{figure}

We evaluate the semiclassical amplitudes including the dilatonic sources
as well as the dilaton propagators.
The dilaton-scalar-scalar vertex (Fig.~\ref{dss}) is given by the
expression
\begin{equation}
V(p',p)=-ia\,(p\cdot p'+m^2\,)\,.
\end{equation}
Since the amplitude of one mediating dilaton,
which is shown in Fig.~\ref{iMd1},
is found to be
\begin{equation}
i\mathcal{M}^{(1)}_d
=-a^2M\, \frac{1}{\vk^2}(EE'-\vp\cdot \vp'+m^2)
\,.
\end{equation}
The lowest-order potential of the dilatonic force can be read from the
amplitude as
\begin{equation}
V_{d1}(r)=-\frac{a^2Mm}{4\pi r}\left(\frac{m}{E}\right)
\approx -\frac{a^2Mm}{4\pi r}\left(1-\frac{\vp^2}{2m^2}\right)\,.
\end{equation}

Next, we obtain the vertices contained in the second-order diagrams as
the following:
\begin{itemize}
\item The dilaton-dilaton-graviton vertex (Fig.~\ref{gdd})
\begin{figure}[ht]
\begin{center}
\includegraphics[width=5cm]{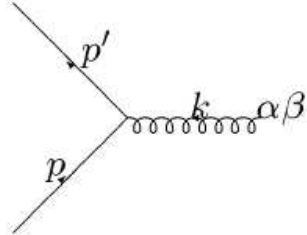}
\caption{Dilaton-dilaton-graviton vertex.}
\label{gdd}
\end{center}
\end{figure}
\begin{equation}
V^{\alpha\beta}(p',p)=-\frac{i\kappa}{2}
[p'^\alpha p^\beta+p'^\beta p^\alpha-\eta^{\alpha\beta}p'\cdot p\,]\,.
\end{equation}

\item The graviton-dilaton-scalar-scalar vertex (Fig.~\ref{gdss})
\begin{figure}[ht]
\begin{center}
\includegraphics[width=6cm]{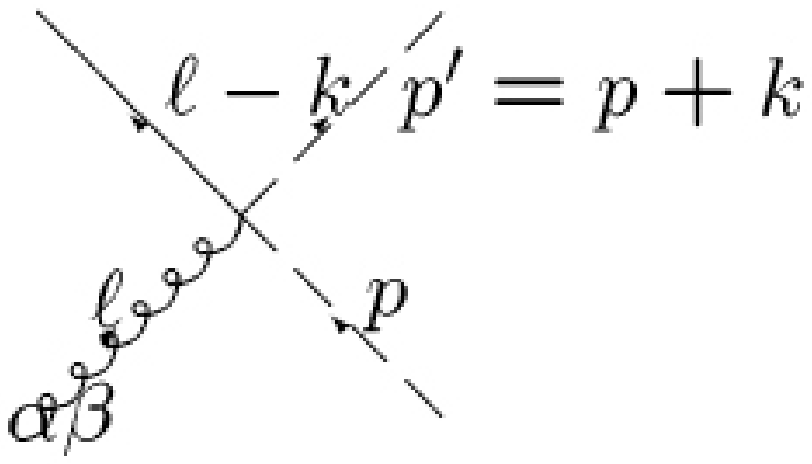}
\caption{Graviton-dilaton-scalar-scalar vertex.}
\label{gdss}
\end{center}
\end{figure}
\begin{equation}
V^{\alpha\beta}(p',p)=\frac{i\kappa a}{2}
[p'^\alpha p^\beta+p'^\beta p^\alpha-\eta^{\alpha\beta}(p'\cdot p
+m^2)\,]\,.
\end{equation}

\item The photon-photon-dilaton vertex (Fig.~\ref{ppd})
\begin{figure}[ht]
\begin{center}
\includegraphics[width=5cm]{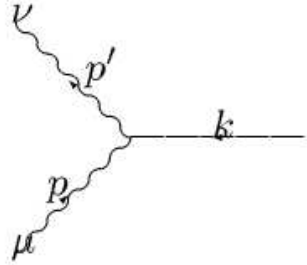}
\caption{Photon-photon-dilaton vertex.}
\label{ppd}
\end{center}
\end{figure}
\begin{equation}
V^{\mu,\nu}(p',p)=i2a \eta^{\mu\nu} p\cdot p'\,.
\end{equation}

\item The photon-dilaton-scalar-scalar vertex (Fig.~\ref{pdss})
\begin{figure}[ht]
\begin{center}
\includegraphics[width=6cm]{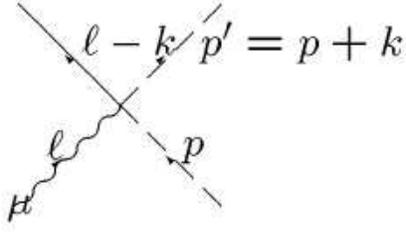}
\caption{Photon-dilaton-scalar-scalar vertex.}
\label{pdss}
\end{center}
\end{figure}
\begin{equation}
V^{\mu}(p',p)=iaq(p^\mu+p'^\mu)\,.
\end{equation}

\item The dilaton-dilaton-scalar-scalar vertex (Fig.~\ref{ddss})
\begin{figure}[ht]
\begin{center}
\includegraphics[width=6cm]{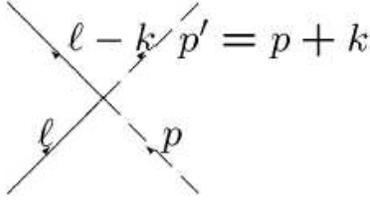}
\caption{Dilaton-dilaton-scalar-scalar vertex.}
\label{ddss}
\end{center}
\end{figure}
\begin{equation}
V(p',p)=ia^2(p\cdot p'-m^2)\,.
\end{equation}

\end{itemize}

Next, we show the amplitudes with diagrams:
\begin{itemize}
\item The amplitude for two dilatonic
external fields, $i\mathcal{M}^{(2\alpha)}$ (Fig.~\ref{2alpha}),
is given by the expression
\begin{figure}[ht]
\begin{center}
\includegraphics[width=6cm]{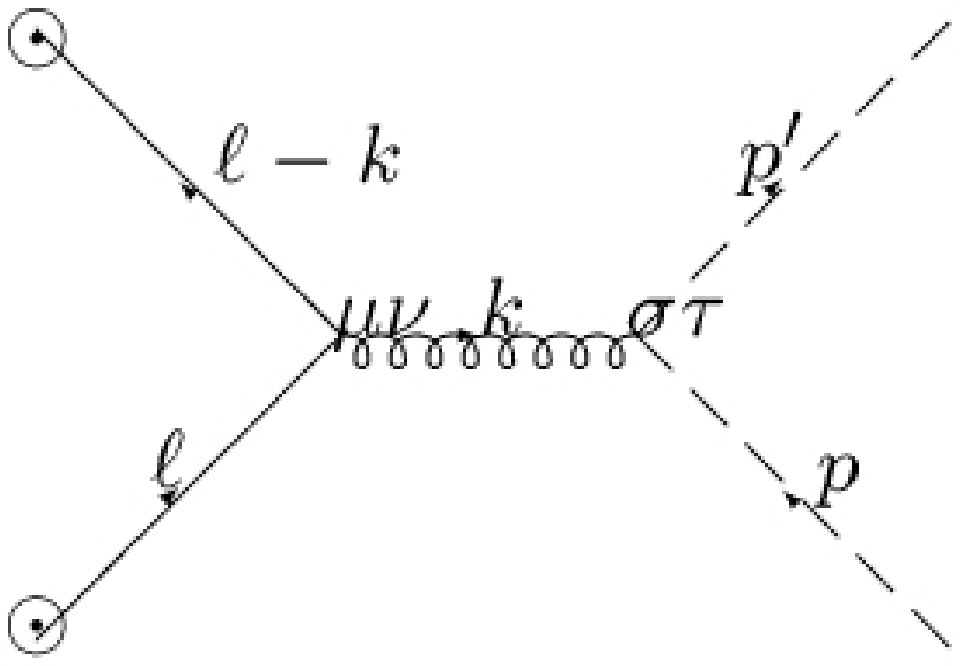}
\caption{Amplitude for two dilatonic
external fields, $i\mathcal{M}^{(2\alpha)}$.}
\label{2alpha}
\end{center}
\end{figure}
\begin{eqnarray}
i\mathcal{M}^{(2\alpha)}&=&\frac{i}{2}\int^\Lambda
\frac{d^3\vl}{(2\pi)^3}\,\phi^{ext}({\bf
l})\,\phi^{ext}({\bf
l-k})V^{\alpha\beta}(\ell-k,\ell)\,
\frac{i\mathcal{P}_{\alpha\beta,\sigma\tau}}{k^2}
V^{\sigma\tau}(p',p)\nonumber \\
&=&-\frac{8\pi Ga^2M^2}{|\vk|^2}\int^\Lambda
\frac{d^3\vl}{(2\pi)^3}\,
\frac{2(\vl-\vk)\cdot\vp'\,\vl\cdot\vp
-\vl\cdot(\vl-\vk)\frac{1}{2}\vk^2}%
{\vl^2(\vl-\vk)^2}
\nonumber \\&=&\frac{8\pi Ga^2M^2}{32|\vk|}
\left(\vp^2-\frac{7}{4}\vk^2\right)+{\cal
O}(\Lambda)\,.
\end{eqnarray}

\item The amplitude for one dilatonic external field and one
gravitational external field, $i\mathcal{M}^{(2\beta)}$
(Fig.~\ref{2beta}), is given by the expression below.
\begin{figure}[ht]
\begin{center}
\includegraphics[width=6cm]{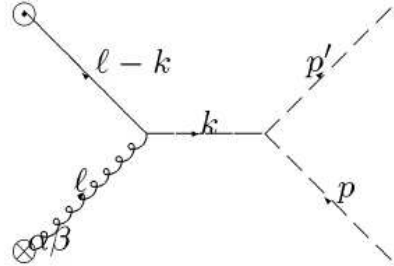}
\caption{Amplitude for one dilatonic external field and one
gravitational external field, $i\mathcal{M}^{(2\beta)}$.}
\label{2beta}
\end{center}
\end{figure}
\begin{equation}
i\mathcal{M}^{(2\beta)}=i\int
\frac{d^3\vl}{(2\pi)^3}\,h_{\alpha\beta}^{ext}({\bf
l})\,\phi^{ext}({\bf l-k})V^{\alpha\beta}(\ell-k,-k)
\frac{i}{k^2}\,
V(p',p)=0\,.
\end{equation}

\item The amplitude for one dilatonic external field and one
gravitational external field, $i\mathcal{M}^{(2\gamma)}$
(Fig.~\ref{2gamma}), is given by the expression
\begin{figure}[ht]
\begin{center}
\includegraphics[width=6cm]{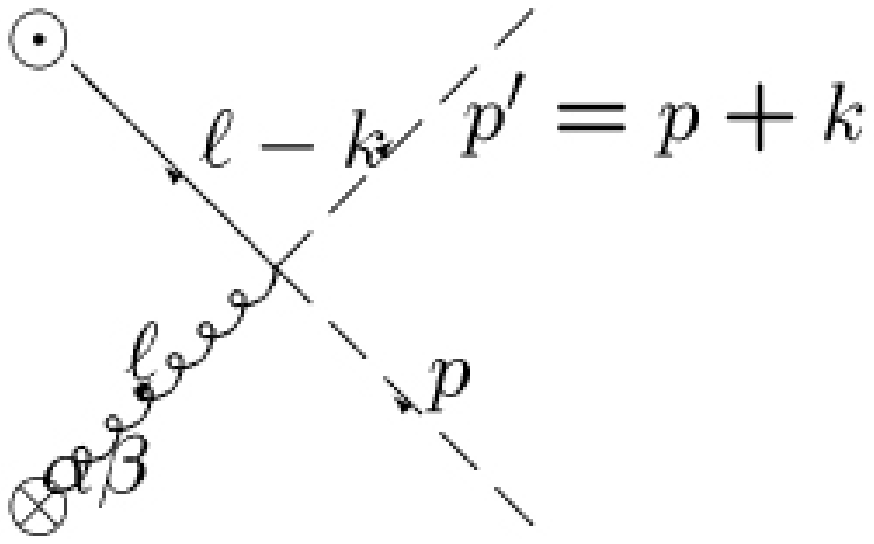}
\caption{Amplitude for one dilatonic external field and one
gravitational external field, $i\mathcal{M}^{(2\gamma)}$}
\label{2gamma}
\end{center}
\end{figure}
\begin{eqnarray}
i\mathcal{M}^{(2\gamma)}&=&i\int
\frac{d^3\vl}{(2\pi)^3}\,\phi^{ext}({\bf
l-k})h_{\alpha\beta}^{ext}({\bf
l})V^{\alpha\beta}(p',p)\nonumber \\&=&-\frac{\kappa^2a^2M^2}{8}
\int\frac{d^3\vl}{(2\pi)^3}
\frac{1}{\vl^2(\vl-\vk)^2}\,(4E^2+2m^2)\nonumber \\
&=&-\frac{\kappa^2a^2M^2}{64|\vk|}\,(6m^2+4\vp^2)=
-\frac{\pi Ga^2M^2}{|\vk|}\,(3m^2+2\vp^2)\,.
\end{eqnarray}

\item The amplitude for one dilatonic external field, one
gravitational external field and an internal scalar,
$i\mathcal{M}^{(2\delta)}$ (Fig.~\ref{2delta}), is given by the
expression below.
\begin{figure}[ht]
\begin{center}
\includegraphics[width=5cm]{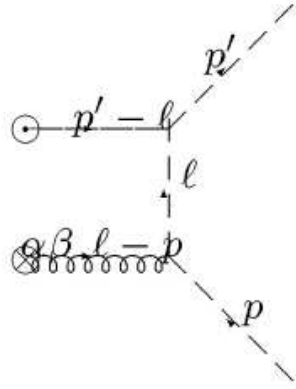}
\caption{Amplitude for one dilatonic external field, one
gravitational external field and an internal scalar,
$i\mathcal{M}^{(2\delta)}$}
\label{2delta}
\end{center}
\end{figure}
\begin{eqnarray}
i\mathcal{M}^{(2\delta)}&=&2i\int\frac{d^3\vl}{(2\pi)^3}\,
\phi^{ext}({\bf
p'-l})V(p',\ell)\frac{i}{\ell^2-m^2+i\epsilon}
V^{\alpha\beta}(\ell,p)h_{\alpha\beta}^{ext}({\bf
l-p})\nonumber \\&=&16\pi GM^2a^2\,{\int\frac{d^3\vl/(2\pi)^3\quad
(2E^2-m^2)\,(E^2-\vl\cdot\vp'+m^2)}{[({\bf p'}-{\bf l})^2+\mu^2][({\bf
p}-{\bf l})^2+\mu^2]({\bf p}^2-{\bf
l}^2+i\epsilon)}}\,.
\end{eqnarray}

\item The amplitude for two electric external
fields creating a dilaton external field, $i\mathcal{M}^{(2\epsilon)}$
(Fig.~\ref{2epsilon}), is given as the expression below.
\begin{figure}[ht]
\begin{center}
\includegraphics[width=7cm]{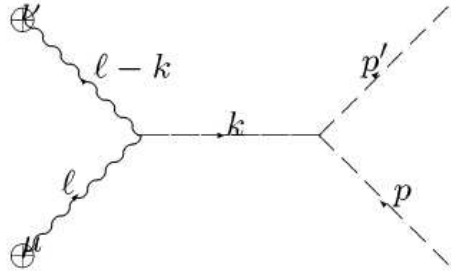}
\caption{Amplitude for two electric external
fields creating a dilaton external field, $i\mathcal{M}^{(2\epsilon)}$.}
\label{2epsilon}
\end{center}
\end{figure}
\begin{eqnarray}
i\mathcal{M}^{(2\epsilon)}&=&\frac{i}{2}\int^\Lambda
\frac{d^3\vl}{(2\pi)^3}A_{\mu}^{ext}({\bf
l})A_{\nu}^{ext}({\bf
l-k})V^{\mu,\nu}(\ell-k,\ell)
\frac{i}{k^2}
V(p',p)\nonumber \\&=&-\frac{a^2Q^2}{\vk^2}\int^\Lambda
\frac{d^3\vl}{(2\pi)^3}\,\frac{\vl\cdot(\vl-\vk)}{\vl^2(\vl-\vk)^2}
\,(p'\cdot p+m^2)\nonumber \\
&=&\frac{a^2Q^2}{16|\vk|}\left(2m^2+\frac{1}{2}\vk^2\right)+{\cal
O}(\Lambda)=\frac{\pi^2
a^2}{|\vk|}\,\frac{Q^2}{(4\pi)^2}\left(2m^2+\frac{1}{2}\vk^2\right)+{\cal
O}(\Lambda)\,.
\end{eqnarray}

\item The amplitude for one dilatonic external
field and one electric external field, $i\mathcal{M}^{(2\zeta)}$
(Fig.~\ref{2zeta}), is given as below.
\begin{figure}[ht]
\begin{center}
\includegraphics[width=6cm]{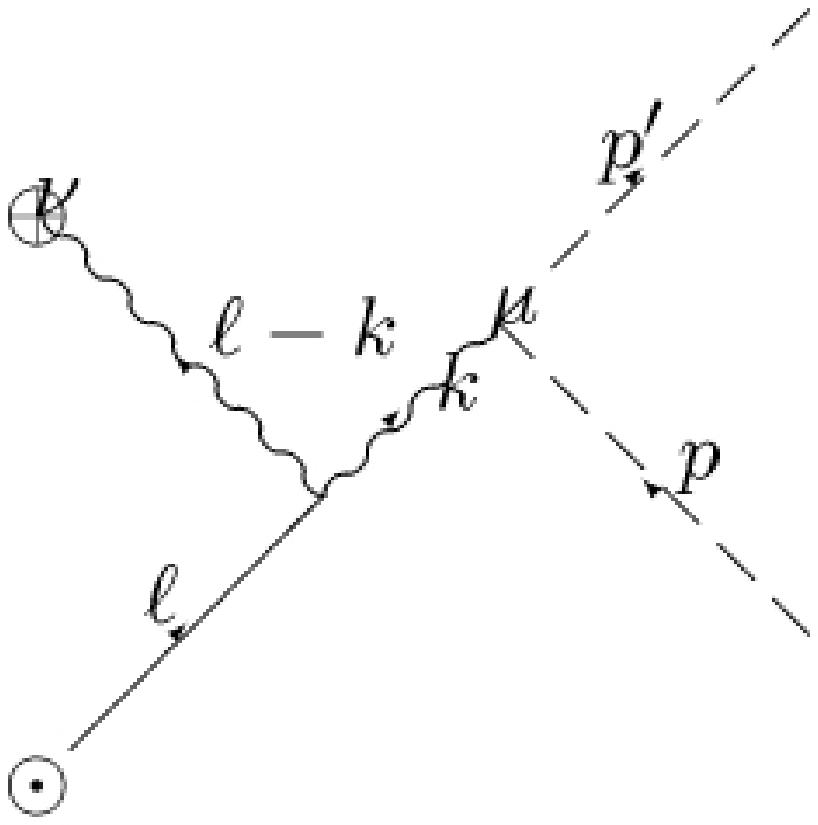}
\caption{Amplitude for one dilatonic external
field and one electric external field, $i\mathcal{M}^{(2\zeta)}$.}
\label{2zeta}
\end{center}
\end{figure}
\begin{eqnarray}
i\mathcal{M}^{(2\zeta)}&=&i\int
\frac{d^3\vl}{(2\pi)^3}A_{\nu}^{ext}({\bf
l-k})\phi^{ext}({\bf
l})V^{\mu,\nu}(\ell-k,-k)
\frac{-i\eta_{\mu\lambda}}{k^2}\,
V^{\lambda}(p',p)\nonumber \\&=&
\frac{2a^2MqQ}{\vk^2}
\int\frac{d^3\vl}{(2\pi)^3}
\frac{\vk\cdot(\vl-\vk)}{\vl^2(\vl-\vk)^2}\,(p^0+p'^0)\nonumber \\
&=&-\frac{a^2MqQ}{8|\vk|}\,(2E)
=-\frac{\pi a^2M}{2}\,\frac{qQ}{4\pi}\,
\frac{2E}{|\vk|}\,.
\end{eqnarray}

\item The amplitude for one dilatonic external
field and one electric external field, $i\mathcal{M}^{(2\eta)}$
(Fig.~\ref{2eta}), is given as
\begin{figure}[ht]
\begin{center}
\includegraphics[width=6cm]{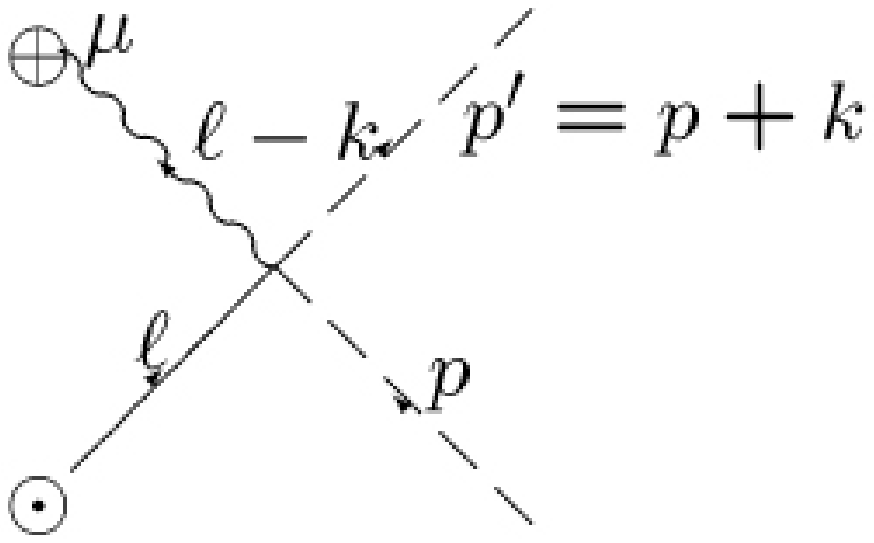}
\caption{Amplitude for one dilatonic external
field and one electric external field, $i\mathcal{M}^{(2\eta)}$.}
\label{2eta}
\end{center}
\end{figure}
\begin{eqnarray}
i\mathcal{M}^{(2\eta)}&=&i\int
\frac{d^3\vl}{(2\pi)^3}A_{\mu}^{ext}({\bf
l-k})\phi^{ext}({\bf
l})V^{\mu}(p',p)\nonumber \\&=&
{a^2MqQ}
\int\frac{d^3\vl}{(2\pi)^3}
\frac{1}{\vl^2(\vl-\vk)^2}\,(p^0+p'^0)\nonumber \\
&=&\frac{a^2MqQ}{8|\vk|}\,(2E)
=\frac{\pi a^2M}{2}\,\frac{qQ}{4\pi}\,
\frac{2E}{|\vk|}\,.
\end{eqnarray}

\item The amplitude for one dilatonic external
field, one electric external field and an
internal scalar, $i\mathcal{M}^{(2\theta)}$ (Fig.~\ref{2theta}),
is given as the expression
\begin{figure}[ht]
\begin{center}
\includegraphics[width=7cm]{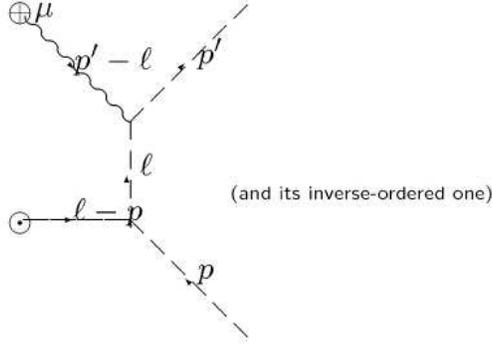}
\caption{Amplitude for one dilatonic external
field, one electric external field and an
internal scalar, $i\mathcal{M}^{(2\theta)}$.}
\label{2theta}
\end{center}
\end{figure}
\begin{eqnarray}
i\mathcal{M}^{(2\theta)}&=&2i\int\frac{d^3\vl}{(2\pi)^3}
A_{\mu}^{ext}({\bf
p'-l})V^{\mu}(p',\ell)\frac{i}{\ell^2-m^2+i\epsilon}
V(\ell,p)\phi^{ext}({\bf
l-p})\nonumber \\&=&
-2a^2MqQ\,{\int\frac{d^3\vl/(2\pi)^3\qquad
(2E)\,(2m^2+\vp^2-\vl\cdot\vp)}{[({\bf p'}-{\bf
l})^2+\mu^2][({\bf p}-{\bf l})^2+\mu^2]({\bf p}^2-{\bf
l}^2+i\epsilon)}}\,.
\end{eqnarray}

\item The amplitude for two dilatonic
external fields, $i\mathcal{M}^{(2\iota)}$ (Fig.~\ref{2iota}),
is given as the expression
\begin{figure}[ht]
\begin{center}
\includegraphics[width=4cm]{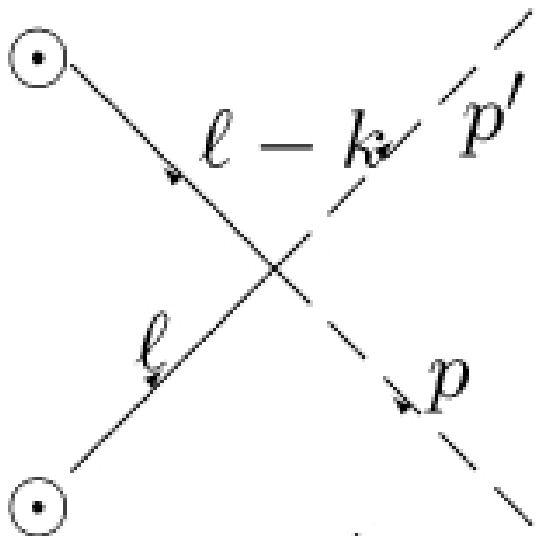}
\caption{Amplitude for two dilatonic
external fields, $i\mathcal{M}^{(2\iota)}$.}
\label{2iota}
\end{center}
\end{figure}
\begin{eqnarray}
i\mathcal{M}^{(2\iota)}&=&
\frac{i}{2}\int\frac{d^3\vl}{(2\pi)^3}\phi^{ext}({\vl})\phi^{ext}({\bf
\vl-k})V(p',p)\nonumber\\&=&
-\frac{a^4M^2}{2}(p\cdot p'-m^2){\int\frac{d^3\vl}{(2\pi)^3}\frac{1}{{\bf
l}^2({\bf l-k})^2}}=-\frac{a^4M^2}{16|{\bf
k}|}\left(\frac{{\bf k}^2}{2}\right)\,.
\end{eqnarray}

\item The amplitude for two dilatonic exernal
fields and an internal scalar, $i\mathcal{M}^{(2\kappa)}$
(Fig.~\ref{2kappa}), is given as
\begin{figure}[ht]
\begin{center}
\includegraphics[width=5cm]{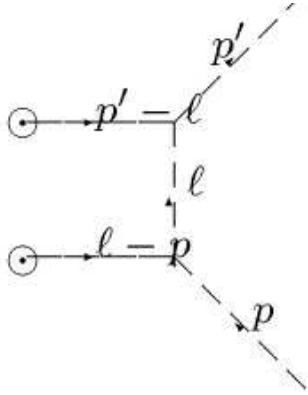}
\caption{Amplitude for two dilatonic exernal
fields and an internal scalar, $i\mathcal{M}^{(2\kappa)}$.}
\label{2kappa}
\end{center}
\end{figure}
\begin{eqnarray}
i\mathcal{M}^{(2\kappa)}&=&i\int\frac{d^3\vl}{(2\pi)^3}
\phi^{ext}({\bf
p'-l})V(p',\ell)\frac{i}{\ell^2-m^2+i\epsilon}
V(\ell,p)\phi^{ext}({\bf
l-p})\nonumber \\&=&
\frac{a^4M^2}{8\pi^3}{\int\frac{d^3\vl \quad
(E^2-\vp\cdot\vl+m^2)(E^2-\vp'\cdot\vl+m^2)}{[({\bf p'}-{\bf
l})^2+\mu^2][({\bf p}-{\bf l})^2+\mu^2]({\bf p}^2-{\bf
l}^2+i\epsilon)}}\,.
\end{eqnarray}

\end{itemize}
We define $\mathcal{M}_d^{(2~total)}$ as the sum of these amplitudes.
Consequently, the second-order potential from the diagrams shown above is
\begin{equation}
V_{d2}(r)=\int\frac{d^3\vk}{(2\pi)^3}e^{i{\bf k}\cdot{\bf
r}}\left[\frac{i\mathcal{M}_d^{(2~total)}}{2E}-\sum_{AB}\int
\frac{d^3\vl}{(2\pi)^3}
\frac{\frac{i\mathcal{M}^{(1)}_A({\bf
p},\vl)}{\sqrt{EE''/M^2}}\frac{i\mathcal{M}^{(1)}_B(\vl,{\bf
p})}{\sqrt{E''E/M^2}}}{E-E''}\right]\,,
\end{equation}
where the sum is calculated over $(A, B)=(G,d), (d,G), (e,d), (d,e),
(d,d)$. Here,
the first-order graviton-mediated amplitude is
given by~\cite{Paszko}
\begin{eqnarray}
i\mathcal{M}_G^{(1)}(\vp',\vp)
&=&ih^{ext}_{\alpha\beta}({\bf k})V^{\alpha\beta}(p',p)\nonumber\\
&=&\frac{4\pi GM}{{\bf
k}^2}(\eta_{\alpha\beta}-2\delta_\alpha^0\delta_\beta^0)
[p'^\alpha p^\beta+p'^\beta p^\alpha-\eta^{\alpha\beta}(p'\cdot p-m^2)]
\nonumber\\
&=&-\frac{4\pi GM}{{\bf k}^2}(4EE'-2m^2)\,,
\label{inelg}
\end{eqnarray}
where $p=(E, \vp)$ and $p'=(E', \vp')$, and
the photon-mediated amplitude at the lowest order
is~\cite{Paszko}
\begin{equation}
i\mathcal{M}^{(1)}_e(\vp',\vp)=
iA^{ext}_{\alpha}({\bf k})V^{\alpha}(p',p)
=\frac{Qq}{{\bf k}^2}\, \eta_{\alpha 0}[p^\alpha+p'^\alpha]
=\frac{Qq}{{\bf k}^2}(E+E')\,.
\end{equation}

The second-order potential can be computed for $\vp=\v0$, and it is found
to be
\begin{equation}
V_{d2}(r)={+\frac{a^2Q^2m}{2(4\pi)^2 r^2}+\frac{a^2M^2m}{(4\pi)^2r^2}
-\frac{a^2MQq}{(4\pi)^2r^2} +\frac{a^4M^2m}{2(4\pi)^2r^2}}\,.
\end{equation}
It is noteworthy that $4\pi G=1$ in this expression.

Finally, we obtain the potential up to $O(1/r^2)$ for gravitational,
electrical, and dilatonic forces. By combining the above result with the
potential obtained in
\cite{Paszko}, and accounting for the interchanging symmetry, we
obtain the static potential of the two particles labeled 
$1$ and $2$, that is
\begin{eqnarray}
V_{Ged}(r)&=&\frac{q_1q_2-(1+a^2)m_1m_2}{4\pi
r}+\frac{(1+a^2)^2m_1m_2(m_1+m_2)}{32\pi^2r^2}\nonumber \\
&&+\frac{(1+a^2)(m_1q_2^2+m_2q_1^2)}{32\pi^2r^2}-
\frac{(1+a^2)q_1q_2(m_1+m_2)}{16\pi^2r^2}\,.
\end{eqnarray}

It is obvious that if  $q_1=\sqrt{1+a^2}m_1$ and
$q_2=\sqrt{1+a^2}m_2$, the potential is completely cancelled for any  
distance. This balance condition is realized in the exact solution in
Ref.~\cite{KSJMP}. 

\section{perihelion precession
\label{precession}}

Thus far, we have derived the next-to-leading classical interactions in
the Einstein-Maxwell system with a dilaton field.
It is of theoretical interest to study the perihelion precession if
there is a binary of charged dilatonic black holes.
The two-body problem of electric charges in general relativity was
considered by Barker and O'Connell~\cite{BO3}.
We attempt to apply our result to their formulae for the perihelion
precession.

Our previous result leads to
the effective Lagrangian for two particles with masses $m_i$ and charges
 $e_i$ $(i=1,2)$ (we use $c=1$) as below
\begin{equation}
{\cal L}'=\frac{1}{2}\mu v^2+\frac{G'\mu
M}{r}+\frac{1}{8}\mu\, k_1 v^4+\frac{3}{2}k_2
\frac{G'\mu Mv^2}{r}-\frac{1}{2}k_3\frac{{G'}^2 \mu M^2}{r^2}\,,
\end{equation}
where $r\equiv |{\bf r}|=|{\bf r}_2-{\bf r}_1|$ (the separation), ${\bf
v}\equiv{\bf v}_2-{\bf v}_1$ (the relative velocity), 
$M\equiv m_1+m_2$ (the total mass), $\mu\equiv m_1m_2/M$ (the reduced
mass) and
\begin{equation}
G'\equiv (1+a^2)G-\frac{e_1e_2}{m_1m_2}\,,
\end{equation}
\begin{equation}
k_1=1-\frac{3\mu}{M}\,,
\end{equation}
\begin{equation}
k_2=\frac{1-a^2/3}{1+a^2}\left(1+\frac{e_1e_2}{G'm_1m_2}\right)+
\frac{2\mu}{3M}\,,
\end{equation}
\begin{equation}
k_3=1+\frac{\mu}{M}
+\frac{e_1^2m_2+e_2^2m_1}{{G'}\mu
M^2}\left(1+\frac{e_1e_2}{G'm_1m_2}\right)-
\left(\frac{e_1e_2}{G'm_1m_2}\right)^2\,.
\end{equation}

It is noteworthy that the coordinate transformation according to Barker
and O'Connell~\cite{BO1,BO2}
\begin{equation}
{\bf r}\rightarrow {\bf r}\left(1+\frac{G'\mu}{2r}\right)\,,
\end{equation}
has been carried out.

The magnitude of precession is given by~\cite{BO3,BO1}, using the above
parameters, as follows:
\begin{equation}
\frac{\left(\frac{1}{2}k_1+3k_2-\frac{1}{2}k_3\right)
G'M\bar{\omega}}{\bar{a}(1-\varepsilon^2)}\,,
\end{equation}
where $\bar{a}$ denotes the semimajor axis, $\varepsilon$ denotes the
eccentricity of the orbit, and $\bar{\omega}$ denotes the average orbital
angular velocity.

It is interesting to see that only $k_2$ depends on the dilaton
parameter $a$ if the parameter of the leading force, $G'$, is fixed. 
Therefore, we find that if other conditions are unchanged and only the
value of
$a$ increases from zero, the precession is reduced.

\section{Second-order external fields
\label{external}}

In this section, we examine the correspondence of the expansion of the
exact solution and the external fields obtained perturbatively when the
balance condition $Q=\sqrt{1+a^2}M$ is satisfied.
We propose that there is one charged source at the origin, and the exact
solution as given by Eqs.~(\ref{Smetric}) and (\ref{Ssol2}) is expressed
using
\begin{equation}
V(r)=1+\frac{(1+a^2) M}{4\pi r}\,.
\end{equation}

The expansions of the metric components of Eq.~(\ref{Smetric}) become
\begin{eqnarray}
g_{00}(r)&=&V^{-\frac{2}{1+a^2}}=1-\frac{2M}{4\pi
r}+\frac{(3+a^2)M^2}{(4\pi r)^2}+\cdots\,,\\
g_{ij}(r)&=&-V^{\frac{2}{1+a^2}}\delta_{ij}=-\left(1+\frac{2M}{4\pi
r}+\frac{(1-a^2)M^2}{(4\pi r)^2}+\cdots\right)\delta_{ij}\,.
\end{eqnarray}
This can be rewritten as
\begin{eqnarray}
g_{00}(r)&=&1-\frac{2M}{4\pi r}+\frac{2M^2}{(4\pi
r)^2}+\frac{Q^2}{(4\pi
r)^2}\cdots\,,\\
g_{ij}(r)&=&-\left(1+\frac{2M}{4\pi
r}+\frac{3M^2}{2(4\pi r)^2}-\frac{Q^2}{2(4\pi
r)^2}-\frac{a^2M^2}{2(4\pi
r)^2}+\cdots\right)\delta_{ij}\,,
\end{eqnarray}
by looking at the amplitudes where the scalar
probe couples to one graviton \cite{Paszko}. In other words, these
amplitudes are interpreted as the interaction of the perturbed external
field and the energy-momentum tensor of the charged dilatonic scalar
field~\cite{Paszko}.

Similarly, the expansion of the electric potential becomes
\begin{equation}
A_0(r)=\frac{1}{\sqrt{1+a^2}}(1-V^{-1})=\frac{\sqrt{1+a^2} M}{4\pi
r}-\frac{(1+a^2)^{3/2} M^2}{(4\pi r)^2}+\cdots\,,
\end{equation}
and this expression can be rewritten as
\begin{equation}
A_0(r)=\frac{Q}{4\pi
r}-\frac{MQ}{(4\pi r)^2}-\frac{a^2MQ}{(4\pi r)^2}+\cdots\,.
\end{equation}
This is due to the amplitudes for which the electric
charge of the scalar probe appears.

Finally, the expansion of the dilaton field becomes
\begin{equation}
\phi(r)=-\frac{1}{2a}\ln V^\frac{2a^2}{1+a^2}
=-\frac{aM}{4\pi r}+\frac{a(1+a^2)M^2}{2(4\pi)^2r^2}+\cdots\,,
\end{equation}
and this expression can be rewritten as
\begin{equation}
\phi(r)=-\frac{aM}{4\pi r}+\frac{aQ^2}{2(4\pi)^2r^2}+\cdots\,.
\label{59}
\end{equation}
Eq.~(\ref{59}) is deduced by the amplitudes
for which the dilaton coupling to the scalar field is considered.

These considerations are mere confirmations of the exact solution in the
second order. However the investigation of this method will be important
if we study the perturbative approach from the exact solution, {\it
i.e.}  the solution with the charge-mass relation which slightly
deviates from the exact balance condition. 

\section{$O(\vp^2)$ Hamiltonian and Lagrangian
\label{Hamiltonian}}

Thus far, we have omitted the momentum-dependent contribution to the
potential in the next-to-leading order. When the leading-order static
potential is cancelled or provides a very small contribution, {\it i.e.}
$Q\approx\sqrt{1+a^2}M$, the
$O(\vp^2/r^2)$ potential cannot be ignored.

The momentum-dependent amplitudes have been shown previously.
We do not show the derivation of the potential again; we directly write
the result here. The Hamiltonian of the probe with
mass $m$ and electric charge $q$ in the system with the fixed mass
$M$ at the origin with the charge $Q$, which is the problem considered in
the present paper, is given by
\begin{eqnarray}
H&=&\frac{{\bf p}^2}{2m}+\frac{(a^2-3)Mm}{2(4\pi)r}\frac{{\bf p}^2}{m}
\nonumber
\\&+&\frac{m}{2(4\pi)^2r^2}\left\{\left(1-\frac{a^2}{2}\right)\left[Q^2-(1+a^2)M^2
\right]+2(3-a^2)M^2\right\}\frac{{\bf
p}^2}{m}
+V(r)\,,
\end{eqnarray}
where
\begin{equation}
V(r)=\frac{Qq-(1+a^2)Mm}{4\pi r}+\frac{(1+a^2)^2M^2m}{2(4\pi)^2
r^2}+\frac{(1+a^2)Q^2m}{2(4\pi)^2 r^2}-\frac{(1+a^2)MQq}{(4\pi)^2
r^2}\,.
\end{equation}
The terms of the order $(1/r)^3$ and higher and the order
$\vp^4$ and higher are neglected.

Subsequently, we obtain the effective Lagrangian for the probe from 
this Hamiltonian as follows:
\begin{eqnarray}
L&=&\frac{m{\bf v}^2}{2}\left[1+\frac{(a^2-3)M}{(4\pi)r}
+\frac{1}{(4\pi)^2r^2}\left\{\left(1-\frac{a^2}{2}\right)\left[Q^2-(1+a^2)M^2
\right]+2(3-a^2)M^2\right\}\right]^{-1}\nonumber \\
& &-V(r)\nonumber \\
&=&\frac{m{\bf v}^2}{2}\left[1-\frac{(a^2-3)M}{(4\pi)r}
-\frac{1}{(4\pi)^2r^2}\left\{\left(1-\frac{a^2}{2}\right)\left[Q^2-(1+a^2)M^2
\right]+(a^2-3)(1-a^2)M^2\right\}\right]\nonumber \\
& &-V(r)\,,
\end{eqnarray}
where ${\bf v}=\vp/m$ and $O(1/r^3)$ is dropped consistently.

In the special case with $Q=\sqrt{1+a^2}M$  and
$q=\sqrt{1+a^2}m$, or in other words, when the static balance condition is
satisfied,%
\footnote{We have already seen that $V$ vanishes in this case.} we obtain
\begin{equation}
L=\frac{m{\bf v}^2}{2}\left[1+\frac{(3-a^2)M}{(4\pi)r}
+\frac{(3-a^2)(1-a^2)M^2}{(4\pi)^2r^2}+O(r^{-3})\right]\,.
\end{equation}
The above result agrees with the classical result up to this
order~\cite{moduliS}.

The investigation of the system of particles with
nearly balanced mass-charge relations by the perturbative method
is of considerable interest.%
\footnote{The study of such systems has thus far been carried out in
Refs.~\cite{SM,KMS}.}

\section{Summary and overview
\label{summary}}

We evaluated the two-body potential of long-range forces coupled with the
dilaton field from the perturbative method associated with the Feynman
diagrams. Subsequently, we verified the correspondences with the known
static exact solutions. Up to the order
$O(1/r^2)$,  we showed the cancellation of the static potential under
the balance condition, $q_i=\sqrt{1+a^2}m_i\quad (i=1,2)$. 

In future, we
wish to examine the higher-order contributions involved in the
cancellation of classical forces.  Further, we wish to study
modified theories such as the higher-derivative theories which
include dimensionful constants; the investigation of the cancellation of
the classical forces are particularly interesting in this case.
Further we want to examine the higher-dimensional
cases, sources with various spins and other extentions of the
perturbative approach.

The calculation of two-body forces on some classical curved backgrounds
may describe many-body problems. The perturbative approach will be
useful in studying such problems.

The perturbative approach is most suitable in accounting for loop
corrections, as in Ref.~\cite{Faller,BB}.
The low-velocity interactions of two or more `particles' are
known to be described in terms of the moduli space associated with them if
the static forces are cancelled.
The structure of the moduli space in the presence of quantum effects is
worth studying, particularly for the case with additional fermion fields
which can control the loop effect.

\begin{acknowledgments}
The authors would like to thank the organizers of JGRG18, where our
partial result ({\tt [arXiv:0902.0412]}) was presented.
We also thank R.~Paszko for information on his related papers.
\end{acknowledgments}


\bibliographystyle{apsrev4-1}

\end{document}